\documentclass{rspublic}

\input BoxedEPS
\SetRokickiEPSFSpecial  
\HideDisplacementBoxes

\newcommand{\be}{\begin{equation}}
\newcommand{\ee}{\end{equation}}
\newcommand{\bea}{\begin{eqnarray}}
\newcommand{\eea}{\end{eqnarray}}
\newcommand{\ben}{\begin{enumerate}}
\newcommand{\een}{\end{enumerate}}
\newcommand{\bit}{\begin{itemize}}
\newcommand{\eit}{\end{itemize}}

\newcommand{\eqn}[1]{equation~(\ref{#1})}

\newcommand{\eqs}[2]{equations~(\ref{#1}) and (\ref{#2})}

\newcommand{\half}{{\textstyle\frac{1}{2}}}

\newcommand\mnl{{\mu\nu\lambda}}

 \newcommand{\perptop}{\genfrac{}{}{0pt}{2}{\perp}{\top}}

\newcommand{\Cite}[1]{(#1)}

\begin{document}

\title[The Theory of the Nucleon Spin]{The Theory of the Nucleon Spin}

\author{R.~L.~Jaffe}

\affiliation{Center for Theoretical Physics and Department of Physics \\
Laboratory for Nuclear Physics
Massachusetts Institute of Technology\\
Cambridge, Massachusetts 02139\\[2ex]
\footnotesize \rm MIT-CTP-3011 \quad hep-ph/0008038\quad August 3,
2000}

\maketitle

\begin{abstract}{transversity, QCD, spin, sum rules}\noindent
I discuss two topics of current interest in the study of the spin 
structure of the nucleon.  First, I discuss whether there is a sum rule 
for the components of the nucleon's angular moments.  Second, 
I discuss the measurement of the nucleon's transversity distribution 
in light of recent results reported by the HERMES collaboration at DESY.
\end{abstract}

\newpage

\section{Introduction}
\label{section1}

Quantum chromodynamics is the only nontrivial quantum field theory
that we are certain describes the real world.  The electroweak part of
the Standard Model, despite its richness, is adequately described by
perturbation theory.  String theory and quantum gravity, despite their
promise, remain far from experimental validation.  Perturbative
methods are applicable to QCD at short distances, where they confirm
that it is the right theory of the strong interactions.  However at
$10^{-13}$ cm, where QCD phenomena are rich and complex, perturbative
methods are hopeless.  Even a simple scattering process like $\pi N
\to \pi\pi N$ at $E_{\rm cm}= 2$ GeV and moderate momentum transfer
seems far beyond our abilities: the energy is too low for perturbative
methods, but too high for chiral dynamics; the process is dominated by
broad overlapping and interfering resonances in the $\pi N$ and
$\pi\pi$ channels; existing lattice methods are useless.

Perhaps QCD will never be solved at this level of detail.  On the
other hand, perhaps it is not important to know every hadron--hadron
scattering amplitude.  Where, then, is it important and interesting to
understand QCD? Each of us has his or her own prejudices.  The
organizers, judging from the program, favor the deep inelastic domain,
where there are two obviously important reasons to study QCD. First,
this is where perturbation theory applies so we can test whether QCD
is the correct theory of the strong interactions.  It has survived all
tests to date, but discovery of even a small deviation would be
extremely exciting.  Second, we need to understand parton distribution
and fragmentation functions in order to use hadron colliders to search
for new physics.  ``QCD Engineering'' as it might be called, is an
essential ingredient in planning the development and exploitation of
the Tevatron and LHC.

Many of us believe there is another important regime where QCD merits
intense study: in the domain of its lowest states, where symmetries
and regularities abound, some of them only poorly understood. 
Examples include the Quark Model, SU(3)-flavor symmetry, the OZI rule,
and vector dominance.  The Quark Model, for example, assumes---for no
good reason---that quark number is conserved in the spectrum of
hadrons.  Thus the nucleon is a $Q^{3}$ state and the pion is $\bar
QQ$.  Huge amounts of spectroscopic data can be cataloged using the
Quark Model and SU(3) flavor symmetry.  However, when we look closely
at hadrons---in deep inelastic scattering experiments, or in
relativistic, field-theoretic models---we find that the nucleon is
full of gluons and $\bar QQ$ pairs, and the pion is often better
regarded as a coherent wave on the chiral condensate as opposed to a
$\bar Q Q$ state.  Why, then, does the simple Quark Model work so
well?  The answer must lie in the confinement dynamics of QCD. This is
an old problem, but the wealth of available data makes low energy QCD
in general, and the quark structure of hadrons in particular, a
tantalizing playground for a theorist with a new idea about
confinement.

There is no uniform framework for theoretical studies of the nucleon
in QCD. We have no relativistic analogue of the Schr\"odinger
wavefunction, which would summarize all that can be asked about the
nucleon in the way it does for the states of the hydrogen atom.  The
best we have at the moment seems to be a list of the expectation
values of local operators in the nucleon state, $\langle
\Theta\rangle_{N}$, and the generalization of this concept to the
parton distribution function, $\langle \Theta(x,Q^{2})\rangle_{N}$.  
Table \ref{tab1} shows a schematic list of local operators.  In the
middle are operators without derivatives in either coordinate or
momentum space.  Upwards, are operators with coordinate space
derivatives.  The simplest ones, those involving derivatives along the
light cone, are given by the moments of parton distributions, $\int dx\,
x^{n}\langle \Theta(x,Q^{2})\rangle_{N}$.  Others, with transverse or
$x^{-}$ derivatives are more complicated (``higher twist'') and harder
to interpret.  Downwards, are operators with insertions of the
coordinate $x^{\mu}$---moments of local operators---equivalent to
derivatives in momentum space.  The classic examples are the magnetic
moment and the mean-square charge radius.  These two can be related to
the derivatives of elastic form factors at zero-momentum transfer. 
Higher moments can be defined, but they {\it are not\/} related to
form factors in a simple way.

A theorist interested in precise and interpretable information about
the nucleon is limited to the expectation values of these few local
operators.  Of course, changing the Dirac and flavor structure and
generalizing to gluonic operators, makes for quite an extensive list
and quite a challenge to our experimentalist colleagues.  But it is
far less than the content of the Schr\"odinger wavefunction.
\begin{table}
\begin{center}
\BoxedEPSF{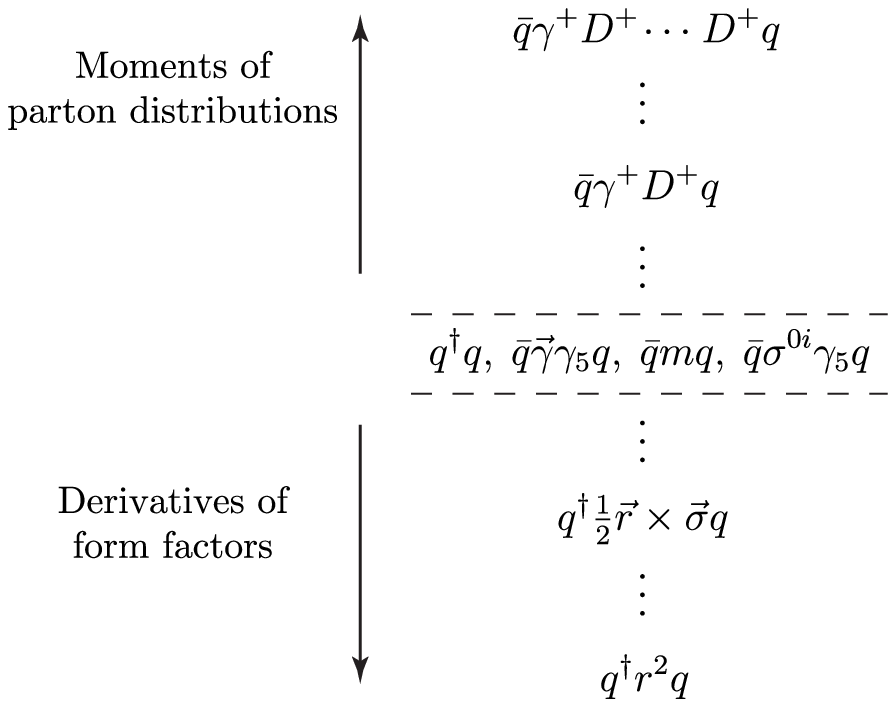}
\end{center}
\caption{}
\label{tab1}
\end{table}
The upward extrapolation of the list in Table
\ref{tab1} leads to the notion of a parton distribution function. 
These give us our most detailed, interpretable, and accessible
description of the nucleon.  At least at leading twist, parton distribution
functions have great heuristic value, providing a snapshot of the
momentum and helicity weighted distributions of quarks and
gluons in a rapidly moving nucleon.  The ``momentum'' variable is
Bjorken's $x$, which is identified with $Q^{2}/2P\cdot Q$ in deep
inelastic scattering and with the parton's momentum fraction in an
infinite momentum frame.  Attempts to generalize this intuitive
picture beyond leading twist and to fragmentation functions have been
only partially successful.

My talk will concern a few topics at the interface between deep
inelastic scattering and hadron structure, where our understanding of
perturbative QCD is so good that we can hope to use it to obtain new
information about the strongly coupled domain where confinement
operates.  Specifically, I want to review what is known about the
measurement of the components of the nucleon's angular momentum and
its transversity.  The situation with angular momentum is not very
satisfactory: we know how to define the $x$-distributions of the
various components of the nucleon's angular momentum, but we do not
know how to measure them.  On the other hand, there is now great
excitement about transversity: Hermes has reported an azimuthal
asymmetry in single particle inclusive deep inelastic scattering,
which, if confirmed, would suggest a large analyzing power for
transversity in a very accessible experiment.

\section{Is there an ``Angular Momentum Sum Rule'' and is it 
experimentally testable?}
\label{section2}

The answers to the questions posed in the title to this section are
``Yes'' and ``Apparently, No'', respectively.  First, I wans to clarify some
terminology.  I would like to distinguish between a sum
rule and an operator relation.  A \emph{sum rule} expresses the expectation
value of a local operator in a state as an \emph{integral} (or sum) over a
distribution measured in an inelastic production process involving the same state. 
This is the traditional definition of a sum~rule, dating back to the Thomas,
Reiche, Kuhn Sum Rule of atomic spectroscopy. All the familiar sum rules
of deep inelastic scattering---Bjorken's, Gross \& Llewellyn-Smith's,
etc.---are this type of relation.  They are even more
powerful because the distribution which is integrated has a simple,
heuristic interpretation as the momentum (Bjorken-$x$) distribution of
the observable associated with the local operator.  The ``spin sum
rule'' gives a typical example:
\begin{eqnarray}
	\langle P,S|\left.\bar q_{a} \gamma^{\mu}\gamma_{5}
	q_{a}\right|_{Q^{2}}|P,S\rangle/S^{\mu} &\equiv&
	\Delta q_{a}(Q^{2})\nonumber\\
	&=& \int_{0}^{1}dx\, \left\{q_{a\uparrow}(x,Q^{2})
	+\bar q_{a\uparrow}(x,Q^{2})\right. \nonumber\\
& &\left.\qquad {}
 -q_{a\downarrow}(x,Q^{2})	-\bar q_{a\downarrow}(x,Q^{2})\right\}
	\label{eq2.0}
\end{eqnarray}
The left-hand side can be measured in $\beta$-decay or other
electroweak processes.  The right-hand side can be measured in deep
inelastic scattering of polarized leptons from polarized targets.  The
meaning of the sum rule is clear because the local operator, $\bar
q_{a} \gamma^{\mu}\gamma_{5} q_{a}$ is the generator of the  
internal rotations (the ``spin'') of the quark field in QCD.  The sum 
rule says the quark's contribution to the nucleon's spin is the 
integral over a spin weighted momentum distribution of the quarks.

Another, less powerful but still interesting type of  relation---sometimes called a
sum rule in the QCD literature---arises simply because an operator can be written
as the sum of two (or more) other operators, $\Theta =
\Theta_{1}+\Theta_{2}$.  If the expectation values of all three
operators can be measured, then this relation, and the assumptions
underlying it, can be tested.  Such a relation exists for the
contributions to the nucleon's angular momentum~\Cite{Jaffe \& Manohar 1990,Ji 1997},
\begin{equation}
	\half = \hat L_{q} + \half\Sigma + \hat J_{g}
	\label{eq2.01}
\end{equation}
where the three terms are {\it roughly\/} the quark orbital angular
momentum, the quark spin, and the total angular momentum on the
gluons.  Ji has shown how, in principal, to measure the various terms in
this relation~\Cite{Ji 1997}.

A sum rule of the classic type also exists for the contributions to
the nucleon's angular
momentum~\Cite{Bashinsky \&
Jaffe 1998, Hagler \& Schafer 1998, Harindranath \& Kundu 1999},
\begin{equation}
	\half = \int_{0}^{1}dx\, \left\{L_{q}(x,Q^{2}) + \half \Delta q(x,Q^{2})
	+ L_{g}(x,Q^{2}) + \Delta G(x, Q^{2})\right\}
	\label{eq2.02}
\end{equation}
where the four terms are {\it precisely\/} the $x$-distributions of
the quark orbital angular momentum, quark spin, gluon orbital angular
momentum, and gluon spin.  However it appears that the
distributions $L_{q}(x,Q^{2})$ and $L_{g}(x,Q^{2})$ are not
experimentally accessible.  So the value of the sum rule is obscure.

Before exploring these relations for the angular momentum in more
depth, let's examine the simpler and well-understood case of energy and 
momentum.

\subsection{Sum rules for energy and momentum}

One hears a lot about the ``momentum sum rule'' in QCD, but nothing
about an ``energy sum rule''.  The reasons are quite instructive. 
Energy and momentum are described by the rank two, symmetric
energy-momentum tensor, $T^{\mu\nu}$,
\begin{equation}
	T^{\mu\nu} = \frac{i}{4}\bar
q(\gamma^{\mu} D^{\nu} +\gamma^{\nu}
	D^{\mu})q +\hbox{h.c.}+ 
	\mathop{\rm Tr}(F^{\mu\alpha}F_{\alpha}^{\nu} - 
	{\textstyle\frac{1}{4}}g^{\mu\nu}F^{2})\, ,
	\label{eq2.1}
\end{equation}
where $D^{\mu}$ and $F^{\mu\nu}$ are the gauge covariant derivative 
and gluon field strength, both matrices in the fundamental representation
of SU(3).\footnote{$T^{\mu\nu}$ is ambiguous up to certain total 
derivatives, but these do not change the arguments presented here.}  

The energy density is given by $T^{00}$,
\begin{equation}
	{\cal E} \equiv T^{00} = \half q^{\dagger}(-i\vec\alpha\cdot\vec D
	+\beta m)q + \hbox{h.c.} + \mathop{\rm Tr}(\vec E^{2}+ \vec B^{2})\, .
	\label{eq2.2}
\end{equation}
The expectation value of $T^{00}$ is normalized,
\begin{equation}
	\langle P | T^{00} | P \rangle = 2E^{2}\, ,
	\label{eq2.3}
\end{equation}
because $|P\rangle$ is an eigenstate of the Hamiltonian, $\int d^{3}x
T^{00}(x)|P\rangle = E|P\rangle$.  This is a good start towards a sum
rule.  However there is no useful sum rule because there is no way to
write any of the terms in \eqn{eq2.2} as an integral over inelastic
production data.  This is not obvious, but the appearance of terms in
${\cal E}$ which are order cubic and higher in the canonical fields is
a bad sign.  The parton distributions of deep inelastic scattering
(DIS) come from operators quadratic in the ``good'' light cone
components of the quark and gluon fields, $q_{+}$ and $\vec
A_{\perp}$~\Cite{Jaffe  1996}. The first term in ${\cal E}$ includes $\bar
q q g$ coupling, and $\vec E^{2}+\vec B^{2}$ involves terms cubic and
quartic in the gluon vector potentials $\vec A_\perp$.

In contrast there is a classic, deep inelastic sum rule for $P^{+}$, 
where $P^{+}=\frac{1}{\sqrt{2}}(P^{0}+P^{3})$, and the 3-direction is 
singled out by the gauge choice $A^{+}=0$.  $T^{++}$ is normalized 
much like $T^{00}$,
\begin{equation}
	\langle P |T^{++}|P\rangle = 2{P^{+}}^{2}.
	\label{eq2.4}
\end{equation}
Unlike $T^{00}$, $T^{++}$ simplifies dramatically in $A^{+}=0$ gauge 
because of the simplification of $D^{+}$ and $F^{+\alpha}$,
\begin{eqnarray}
	D^{+}&=&\partial^{+}-igA^{+} \to \partial^{+}\nonumber\\
	F^{+\alpha}&=&\partial^{+}A^{\alpha}-\partial^{\alpha}A^{+}
	+g[A^{+},A^{\alpha}] \to \partial^{+}A^{\alpha}\, .\label{eq2.5}
\end{eqnarray}
As a result $T^{++}$ is quadratic in the fundamental dynamical
variables, $q_{+}$ and $\vec A_{\perp}$ and all interactions disappear,
\begin{equation}
	T^{++} = iq_{+}^{\dagger}\partial^{+} q_{+} + 
	\mathop{\rm Tr}(\partial^{+}\vec A_{\perp})^{2}\, .
	\label{eq2.6}
\end{equation}
The two terms give the contributions of quarks and gluons respectively 
to the total $P^{+}$.  It is straightforward to relate each to 
an integral over a positive definite parton ``momentum'' distribution,
\begin{eqnarray}
	iq_{+}^{\dagger}\partial^{+} q &\to &\int dx\, x q(x)\nonumber\\
	(\partial^{+}\vec A_{\perp})^{2} &\to &\int dx\, x g(x)\, 
	\label{eq2.7}
\end{eqnarray}
in which the parton probability density is weighted by the 
observable (in this case $x$) appropriate to the sum rule.
Keeping track of renormalization scale dependence and kinematic
factors of $P^{+}$, one obtains the standard ``Momentum'' Sum Rule,
\begin{equation}
	1 = \int_{0}^{1} dx\, x \left\{q(x,Q^{2}) + g(x,Q^{2})\right\}
	\label{eq2.8}
\end{equation}

The lessons learned from this exercise generalize to the more 
difficult case of angular momentum:
\begin{itemize}
	\item The time-components of the tensor densities associated with
	space time symmetries do not yield sum rules. 
	Interactions do not drop out.  They yield relations which are
	difficult to interpret because quark and gluon contributions do
	not separate.  Individual terms are not related to integrals over
	parton distributions.

	\item  The $+$-components of the same tensor densities do yield 
	useful sum rules, which have a parton interpretation in $A^{+}=0$ 
	gauge.  Interactions drop out.  Each term can be represented as an 
	integral over a parton distribution weighted by the 
	appropriate observable quantity. 
\end{itemize}

\subsection{Sum rules for angular momentum}

The situation for angular momentum is not satisfactory.  The
time-component analysis yields a relation, some of whose ingredients
can be measured (in principle) in deeply virtual Compton scattering. 
But it has no place for a separately gauge invariant gluon spin and
orbital angular momentum, no clean separation between quark and gluon
contributions, and no relation to quark or gluon $x$ distributions. 
The $+$ component analysis yields a classic sum rule with separate
quark and gluon spin and orbital angular momentum contributions, each
gauge invariant, each related to a parton distribution and each free
from interaction terms.  Unfortunately, there does not seem to be a
way to measure the terms in this otherwise perfectly satisfactory the
sum rule.

The tensor density associated with rotations and boosts is a three
component tensor antisymmetric in the last two indices, $M^{\mnl}$. 
To extract a sum rule, we polarize the nucleon along the 3-direction
in its rest frame and set $\nu=1,\lambda=2$ in order to select
rotations about this direction.  The matrix elements of $M^{012}$ and
$M^{+12}$ are both normalized in terms of the nucleon's momentum 
($P^{\mu}=(M,0,0,0)$) and spin ($S^{\mu}=(0,0,0,M)$)~\Cite{Jaffe \& Manohar 1990}.

First consider the time component, $M^{012}$  (Ji 1997), 
\begin{equation}
	M^{012} = \frac{i}{2}q^{\dagger}(\vec x \times \vec D)^{3}q
	+\half q^{\dagger}\sigma^{3} q +2 \mathop{\rm Tr} E^{j}
	(\vec x \times  i\vec D)^{3} A^{j} + \mathop{\rm Tr} (\vec E\times
	\vec A)^{3}\, .
	\label{eq2.9}
\end{equation}
The four terms look like the generators of rotations (about the
3-axis) for quark orbital, quark spin, gluon orbital, and gluon spin
angular momentum respectively.  Taking the matrix element in a nucleon
state at rest, one obtains
\begin{equation}
	\half = \hat L_{q} + \half\Sigma + \hat L_{g} + \Delta \hat G\, .
	\label{eq2.10}
\end{equation}
There are problems, however.  There are no parton representations for
$\hat L_{g}$, $\hat L_{q}$, or $\Delta\hat G$, so it is not a sum rule
in the classic sense.  $\Sigma$ is the integral of the helicity
weighted quark distribution, but $\Delta\hat G$ is not the integral of
the helicity weighted gluon distribution.  Interactions prevent a
clean separation into quark and gluon contributions as they did for
$T^{00}$.  And worse still, $\hat L_{g}$ and $\Delta\hat G$ are not
separately gauge invariant, so only the sum $\hat J_{g}=\hat
L_{g}+\Delta\hat G$ is physically meaningful.

The most important feature of the relation, \eqn{eq2.10}, is the result
derived by Ji, that $\hat J_{q}=\hat L_{q} +\half\Sigma$ and $\hat J_{g}$ can,
in principle, be measured in deeply virtual Compton scattering~\Cite{Ji 1997}.

Turning to the $+$\,-component sum rule, we find a much simpler form,
\begin{equation}
	M^{+12} = \half
	q^{\dagger}_{+}(\vec x\times\vec i\partial)^{3}q_{+}
	+\half q_{+}^{\dagger}\gamma_{5} q_{+}
	+ 2\mathop{\rm Tr}F^{+j}(\vec x\times\vec i  \partial)A^{j}
	+\mathop{\rm Tr} \epsilon^{+-ij}F^{+i}A^{j}
	\label{eq2.11}
\end{equation}
in $A^{+}=0$ gauge.\footnote{This gauge condition must be supplemented
by the additional condition that the gauge fields vanish fast enough
at infinity.} The four terms in $M^{+12}$ correspond respectively to
quark orbital angular momentum, quark spin, gluon orbital angular
momentum, and gluon spin, all about the 3-axis.  Each is separately
gauge invariant\footnote{Note however, that in any gauge other than
$A^{+}=0$, the operators are nonlocal and appear to be interaction
dependent.  The same happens to the simple operators involved in the
momentum sum rule, \eqn{eq2.6}} and involves only the ``good'', i.e., 
dynamically independent, degrees of freedom, $q_{+}$ and $\vec
A_{\perp}$.  Each is a generator of the appropriate symmetry
transformation in light-front field theory.  The resulting sum rule,
\begin{equation}
	\half = L_{q} + \half \Sigma + L_{g} + \Delta G
	\label{eq2.12}
\end{equation}
is a classic deep inelastic sum rule.  It can be written as an 
integral over $x$-distributions
\begin{equation}
	\half = \int_{0}^{1}dx\,\left\{L_{q}(x,Q^{2}) + \half \Delta q(x,Q^{2})
	+ L_{g}(x,Q^{2}) + \Delta G(x, Q^{2})\right\}
	\label{eq2.13}
\end{equation}
where each term is an interaction independent, gauge invariant, 
integral over a partonic density associated with the appropriate 
symmetry generator.  

Satisfying though \eqs{eq2.12}{eq2.13} may be from a theoretical point 
of view, they are quite useless unless someone finds a way to measure 
the two new terms $L_{q}$ and $L_{g}$.

\section{The Parton Distribution for Transversity}

There are many different ways of looking at parton distributions that help us 
understand different aspects of their physical significance.  
The transversity is most easily defined in the old fashioned parton
model at infinite momentum, but its properties become clearer if we
also look at it in the language of helicity amplitudes.  First, in the
parton model: consider a nucleon moving with (infinite) momentum in
the $\hat e_{3}$-direction, but polarized along one of the directions
transverse to $\hat e_{3}$.  The transversity, $\delta
q_{a}(x,Q^{2})$, counts the quarks of (flavor $a$ and) momentum
fraction $x$ polarized parallel to the nucleon minus the
number antiparallel.  Together, the unpolarized quark distribution,
$q_{a}(x,Q^{2})$, the quark helicity distribution, $\Delta
q_{a}(x,Q^{2})$, and the transversity provide a complete description
of the quark spin in deep inelastic processes at leading
twist~\Cite{Jaffe  1996}. If quarks moved nonrelativistically in the
nucleon, $\delta q$ and $\Delta q$ would be identical, since rotations
and Euclidean boosts commute and a series of boosts and rotations can
convert a longitudinally polarized nucleon into a transversely
polarized nucleon at infinite momentum.  So the difference between
the transversity and helicity distributions reflects the relativistic
character of quark motion in the nucleon.  There are other important
differences between transversity and helicity.  For example, quark and
gluon helicity distributions ($\Delta q$ and $\Delta g$) mix under
$Q^{2}$-evolution.  There is no gluon analog of  transversity in the
nucleon, so $\delta q$ evolves without mixing, like a ``non-singlet''
distribution function.

\begin{figure}[t]
\begin{center}
\BoxedEPSF{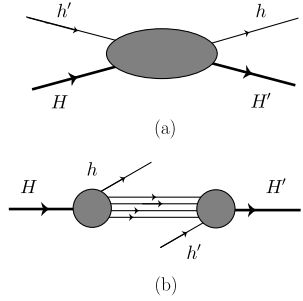 scaled 400} 
\end{center}    
 \caption{Quark hadron forward scattering.  Quark helicities are 
 labelled $h$ and $h'$; hadron helicities are $H$ and $H'$.  (a)~Full 
 scattering amplitude; (b)~u-channel discontinuity which gives the 
 quark distribution function in DIS.}
\label{fig1}
 \end{figure}

Transversity is set apart from other parton distributions by its
chiral transformation properties.  A leading twist quark distribution
can be viewed as a discontinuity in a quark-nucleon forward scattering
amplitude labelled by the helicities of the external lines.  This is
the lower part of the standard ``handbag diagram'' shown in  figure~\ref{fig1}(a).
 
Conservation of angular momentum, parity and time
reversal leave only three independent helicity amplitudes, ${\cal
A}_{++,++}$, ${\cal A}_{+-,+-}$, and ${\cal A}_{+-,-+}$, where the
subscripts denote the quark and nucleon helicities as labelled in the
figure.  At leading twist, quark helicity and chirality are identical. 
The spin average ($q$) and helicity ($\Delta q$) distributions involve
${\cal A}_{++,++}$, ${\cal A}_{+-,+-}$, which preserve quark helicity,
but the transversity corresponds to helicity (and therefore chirality)
flip, ${\cal A}_{+-,-+}$.  This is a simple consequence of quantum
mechanics.  The two states of transverse polarization can be written
as superpositions of helicity eigenstates: 
 $|{\perptop}\rangle= \frac{1}{\sqrt{2}}(|+\rangle\pm|-\rangle)$;
the cross section with transverse polarization has the form
$d\sigma_{\perptop}\propto \langle {\perptop}|\cdots|
{\perptop}\rangle$; so the difference of cross sections is
proportional to helicity flip, $d\sigma_{\perp}-d\sigma_{\top} \propto
\langle +|\cdots|-\rangle + \langle -|\cdots|+\rangle$.  For this
reason, the transversity distribution called ``chiral-odd'', in
contrast to the ``chiral-even'' distributions, $q$ and $\Delta q$.

Quark chirality is conserved at all QCD and electroweak
vertices, however quark chirality can flip in distribution functions 
because they probe the soft regime where chiral symmetry is 
dynamically broken in QCD.  This is another reason to be interested in 
transversity---it probes dynamical chiral symmetry breaking, an 
incompletely understood aspect of QCD.

\begin{figure}
$$\BoxedEPSF{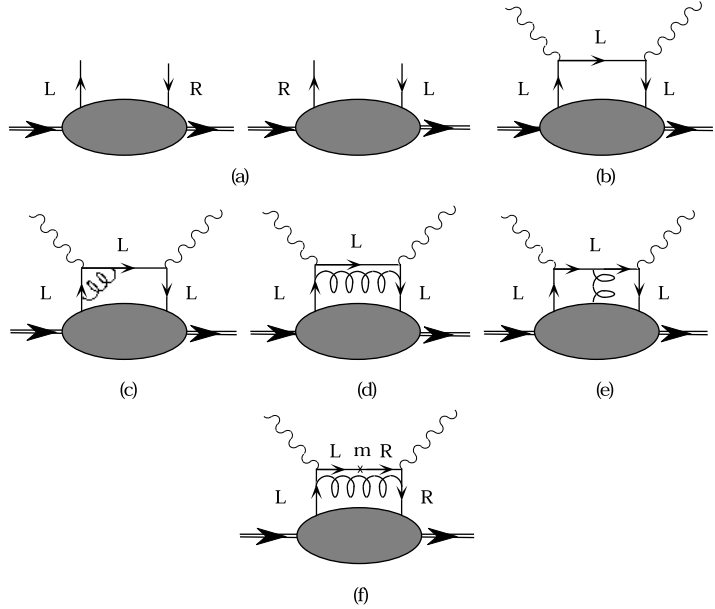 scaled 500}$$
	\caption{{Chirality in deep inelastic scattering:
	(a) Chirally odd contributions to $h_1(x)$; (b)--(e) chirally
	even contributions to deep inelastic 
        scattering (plus $L \leftrightarrow R$
	for electromagnetic currents); (f) chirality flip by mass insertion. }}
	\label{fig2}
\end{figure}

\begin{figure}
$$\BoxedEPSF{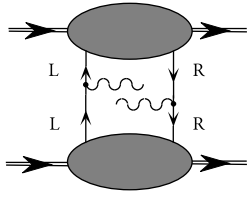 scaled 500}$$
	\caption{{Chirality in Drell-Yan 
	(plus $L \leftrightarrow R$) production of lepton pairs.}}
	\label{fig3}
\end{figure}

Because all hard QCD and electroweak processes preserve chirality,
transversity is difficult to measure.  It decouples from inclusive DIS
and most other familiar deep inelastic processes.  The argument is
made graphically in figure~\ref{fig2}.  
This makes it difficult to observe
transversity.  Some process must flip the quark chirality a
second time.  The classic example, where transversity was discovered
by Ralston and Soper, is transversely polarized Drell-Yan production
of muon pairs: $\vec p_{\perp}\vec p_{\perp}\to \mu^{+}\mu^{-}X$, 
which is shown
diagrammatically in figure~\ref{fig3}.  
Chirality is flipped in both soft
distribution functions and the cross section is proportional to
$\delta q(x_{1},Q^{2})\times\delta\bar q(x_{2},Q^{2})$.

Transversity would not decouple from deep inelastic scattering if some
electroweak vertex would flip chirality.  Unfortunately (and
accidentally from the point of view of QCD) all photon, $W^{\pm}$ and
$Z^{0}$ couplings preserve chirality.  Quark-Higgs couplings
violate chirality but are too weak to be of interest.  Quark mass
insertions flip chirality (see  figure~\ref{fig1}(f)), and indeed a
careful analysis reveals effects proportional to $m\delta
q(x,Q^{2})/\sqrt{Q^{2}}$ in inclusive DIS with a transversely
polarized target. However the $u$, $d$, and $s$ quarks, which are
common in the nucleon, are too light to give significant sensitivity
to $\delta q$.

\begin{figure} 
$$\BoxedEPSF{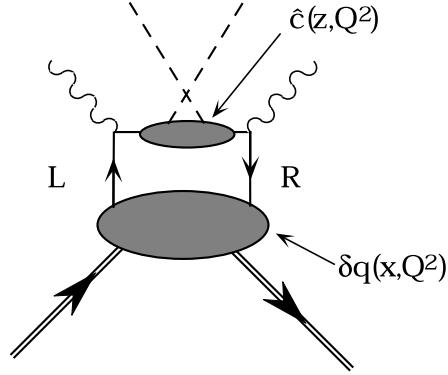 scaled 800}$$
	\caption{{Single particle inclusive scattering $ep\rightarrow ehX$. 
	The chiral-odd fragmentation function, denoted $\hat c$ in the figure, 
	compensates for the chirality flip in the distribution function, 
	denoted $\delta q$ in the figure.}}
	\label{fig4}
\end{figure}

What is needed is an insertion that flips chirality without
introducing a $1/\sqrt{Q^{2}}$ suppression.  Fortunately there are
several candidates in the form of chiral-odd {\it fragmentation
functions\/} which describe the production of hadrons in the current
fragmentation function of deep inelastic scattering.  A generic
example is shown in figure~\ref{fig4}.
Several candidates have been proposed 
and studied in some detail.  
\begin{itemize}
	\item $\delta \hat q_{a}(z,Q^{2})$, the transverse, spin-dependent
	fragmentation function.  This is the analog in fragmentation of
	transversity, and describes the fragmentation of a transversely
	polarized quark into a transversely polarized hadron with momentum
	fraction $z$~\Cite{Jaffe  \& Ji 1993, Boer 2000}. To access $\delta \hat q$, it is
	necessary to measure the spin of a particle in the final state of
	DIS. In practice this limits the application to production of a
	$\Lambda$ hyperon---the only particle whose spin is easy to
	measure through its parity violating decay.
	
	\item $\delta \hat q_{I}(z,m^{2},Q^{2})$, the two pion
	interference fragmentation function~\Cite{Collins et~al.\ 1994, Collins \&
Ladinsky 1994, Jaffe  et al.\ 1998}. This describes
	the fragmentation of a transversely polarized quark into a pair of
	pions whose orbital angular momentum is correlated with the quark
	spin.  This requires measurement of two pions in the final state. 
	It may be quite useful, especially in polarized collider
	experiments.  I will not discuss it further here.
	
	\item $\hat c(z,Q^{2})$, the single particle azimuthal asymmetry
	fragmentation function.  This function, first discussed by
	Collins  et al.~\Cite{1994}. describes the azimuthal
	distribution of pions about the axis defined by the struck quark's
	momentum in deep inelastic scattering.
\end{itemize}
All three of these fragmentation functions are chiral-odd and
therefore produce experimental signatures sensitive to the
transversity distribution in the target nucleon.  Each may play an
important role in future experiments aimed at probing the nucleon's
transversity.  Recently HERMES has announced observation of a spin
asymmetry which seems to be associated with the Collins function,
$c(z,Q^{2})$.  So although all three deserve discussion, I will spend
the rest of my time on the Collins function and the HERMES asymmetry.

\subsection{The Collins Fragmentation Function}

The standard description of fragmentation without polarization
requires a single fragmentation function usually called $D_{h}(z)$. 
It gives the probability that a quark will fragment into a hadron,
$h$, with longitudinal momentum fraction $z$.\footnote{For simplicity
I suppress the dependence of $D$ on the virtuality scale, $Q^{2}$ and
the quark flavor label $a$.} The transverse momentum of $h$ relative
to the quark is integrated out.  If the transverse momentum, $\vec
p_{\perp}$, is observed, then it is possible to construct
distributions weighted by geometric factors.  For instance,
\begin{eqnarray}
	c(z) &\propto & \int d^{2}p_{\perp} D_{h}(z,\vec p_{\perp}) \cos\chi\, , \nonumber \\ 
	\noalign {\hbox{where, for comparison,}}
	D(z) &\propto & \int d^{2}p_{\perp}D_{h}(z,\vec p_{\perp})\, ,
	\label{eq3.1}
\end{eqnarray}
Here $D_{h}(z,\vec p_{\perp})$ is the probability for the quark to
fragment into hadron $h$ with momentum fraction $z$ and transverse
momentum $\vec p_{\perp}$.  $\chi$ is the angle between $\vec
p_{\perp}$ and some vector, $\vec w$, defined by the initial state. 
Since we don't know the direction of the quark's momentum exactly,
the transverse momentum of the hadron, $\vec p_{\perp}$, is
defined relative to some large, externally determined momentum, such 
as the momentum of the virtual photon, $\vec q$, in DIS.

\begin{figure}[htb]
$$\BoxedEPSF{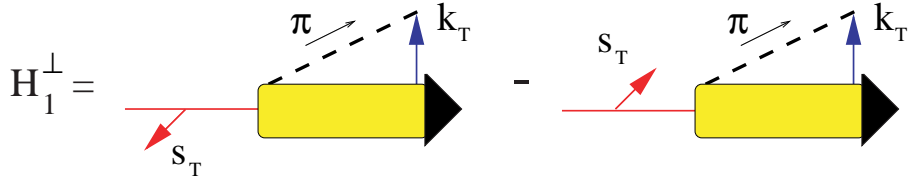 scaled 700}$$
\caption{\label{H1perp}The Collins effect function
$H_1^\perp$ signals different probabilities for $q(\pm \vec S_T) \to
\pi(\vec k_T) + X$.}
\label{fig5}
\end{figure}

How can c(z) figure in deep inelastic scattering?  The trick is to
find a vector, $\vec w$, relative to which $\chi$ can be defined.  If
the target is polarized, it is possible to define $\vec w$ by taking
the cross product of the target spin, $\vec s$, with either the
initial or final electron's momentum ($\vec k$ or $\vec k'$) depending
on the circumstances.  Generically, then, the observable associated
with $c(z)$ is $\cos\chi\propto \vec k \times \vec s \cdot \vec p$,
where $\vec p$ is the momentum of the observed hadron in the final
state.  The situation is illustrated in  figure~\ref{fig5} from 
Boer~\Cite{1999}.
This observable is even under parity (because $\vec s$ is a
pseudovector), but odd under time-reversal.  This \emph{does not}
mean that it violates time reversal invariance.  Instead it means that
it will vanish unless there are final state interactions capable of
generating a nontrivial phase in the DIS amplitude.  This
subtlety makes it hard to find a good model to estimate $c(z)$ because
typical fragmentation models involve only tree graphs (if they involve
quantum mechanics at all!)\  which are real.

The Collins fragmentation function, $c(z)$, may be interesting in 
itself, but it is much more interesting because it is chiral-odd and 
combines with the transversity distribution in the initial nucleon to 
produce an experimentally observable asymmetry sensitive to 
the transversity.  Two specific cases figure in recent and 
soon-to-be-performed experiments.

\subsubsection {Single particle inclusive DIS with a transversely
polarized target: $e \vec p_{\perp}\to e' \pi X$}

If the target is transversely polarized (with respect to the initial
electron momentum, $\vec k$), then $\vec w = \vec k\times\vec s$
defines a vector normal to the plane defined by the beam and the
target spin.  The transverse momentum of the produced hadron can be
defined either with respect to the beam or the momentum transfer $\vec
q$---the difference in higher order in $1/Q$.  $\cos\chi$ is defined
by $\cos \chi= \vec p_{\perp}\cdot\vec w/|\vec p_{\perp}||\vec w|$. 
The kinematics are particularly simple in this case (transverse spin). 
Experimenters prefer to think of the effect in terms of the angle
($\phi$) between two planes: Plane 1 is defined by the virtual photon
and the target spin, and Plane 2 is defined by the virtual photon
and the transverse momentum of the produced hadron.  Then
$\sin\phi=\cos\chi$ and the effect is known as a ``$\sin\phi$''
asymmetry.  When the cross section is weighted by $\sin\phi$ the
result is 
\begin{equation}
	\frac{d\Delta\sigma_{\perp}}{dx\,dy\,dz} = 
	\frac{2\alpha^{2}}{Q^{2}} \sum_{a}e_{a}^{2}\delta 
	q_{a}(x)c_{a}(z) 
	\label{eq3.2}
\end{equation}
where $y=E-E'/E$, and $\Delta\sigma$ is the difference of cross
sections with target spin reversed.\footnote{In principle this
reversal is superfluous because the $\sin\phi$ asymmetry must be odd
under reversal and the rest of the cross section must be even. 
However, it helps experimenters to reduce systematic errors.} This is
a leading twist effect, which scales (modulo logarithms of $Q^{2}$) in
the deep inelastic limit.  If $c(z)$ is not too small, it is will
become the ``classic'' way to measure the nucleon's transversity
distributions.

 No experimental group has yet measured hadron production in deep
 inelastic scattering from a transversely polarized target, so there
 is no data on $\Delta \sigma_{\perp}$.  HERMES at DESY intend to take
 data under these conditions in the next run.  One reason for this was
 the observation of a $\sin\phi$ asymmetry with a {\it
 longitudinally\/} polarized target which HERMES announced last
 year~\Cite{Airapetian 1999}. It strongly suggests, but does not
 require, that $\Delta\sigma_{\perp}$ should be large.

\subsubsection {Single particle inclusive DIS with a longitudinally
polarized target: $e \vec p_{\|}\to e' \pi X$}

\begin{figure}[ht]
  \begin{center}
$$\BoxedEPSF{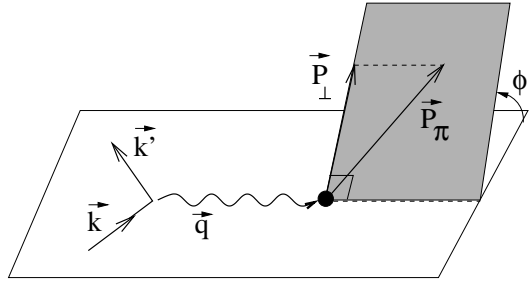 scaled 500}$$
   \end{center}
\begin{minipage}[r]{\linewidth}
   \caption{Kinematic planes for pion production in semi-inclusive deep-inelastic scattering.}
 \label{fig6}
\end{minipage}
 \end{figure}

The possibility of a $\sin\phi$ asymmetry is more subtle in this case
and escaped theorists attention for a long time.  Such an asymmetry was first
pointed out by Oganessian  et al.\Cite{1998}.  As $Q^{2}$ and $\nu$ go to
$\infty$, the initial and final electron's momenta become parallel.  If the
target spin is parallel to $\vec k$, then it is impossible to
construct a vector from $\vec k$ or $\vec k'$ and $\vec s$ in this
limit.  However $\vec k$ and $\vec k'$ are not exactly parallel, so
$\vec s$ has a small component perpendicular to the virtual photon's
momentum, $\vec q = \vec k - \vec k'$.  The vector, $\vec w$, can be
defined as $\vec w = \vec k'\times\vec s$, and the kinematic situation
is shown in  figure~\ref{fig6} from Airapetian~\Cite{1999}.
This produces an asymmetry similar to the previous case, but weighted
by $|\vec s_{\perp}| \propto 2Mx/Q$.  Because this leading (twist two)
effect is kinematically suppressed by $1/Q$, it is necessary to
consider other, twist-three, effects which might be competitive.  A
careful analysis turns up a variety of twist-three effects, leading to
a cross section of the form~\Cite{Boer 1999, Oganessian  et al.\ 1998},
\begin{equation}
	\frac{d\Delta\sigma_{\|}}{dx\,dy\,dz} = 
	\frac{2\alpha^{2}}{Q^{2}} \frac{2Mx}{Q}\sqrt{1-y}\sum_{a}e_{a}^{2}\left\{\delta 
	q_{a}(x)c_{a}(z) + \frac{2-y}{1-y} h_{La}(x)c_{a}(z)\right\}\,  ,
	\label{eq3.3}
\end{equation}
where $h_{L}(x)$ is a longitudinal spin dependent, twist three 
distribution function analogous to $g_{T}$.

\begin{figure}[t]
\begin{center}
\BoxedEPSF{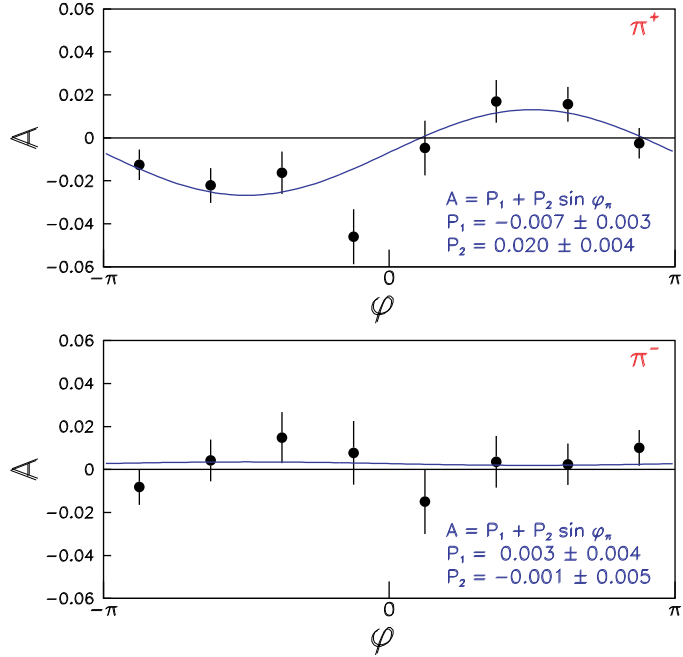 scaled 500} 
\end{center}
 \caption{Azimuthal asymmetry ($\sin\phi$) distribution for $\pi+$ 
 and $\pi^{-}$ production with a longitudinally polarized target at 
 HERMES.}
\label{fig7}
 \end{figure}

By far the most interesting thing about $\Delta \sigma_{\|}$ is that
HERMES has seen such an asymmetry in their $\pi^{+}$ data.  The HERMES 
data is shown in figure~\ref{fig7}.  
 They see no
effect in their $\pi^{-}$ data.  Because $u$ quarks predominate in the
nucleon, because $e_{u}^{2}=4e_{d}^{2}$, and because $u\to\pi^{+}\gg
u\to\pi^{-}$, they expect no signal in $\pi^{-}$.  They have not
reported on $\pi^{0}$, where an asymmetry similar to $\pi^{+}$ would
be expected~\Cite{K.~A.~Oganessian 2000, personal communication}.

If the HERMES result is confirmed, it demonstrates that the Collins
fragmentation function is nonzero.  Somehow the final state
interactions between the observed pion and the other fragments of the
nucleon suffice to generate a phase that survives the sum over the
other unobserved hadrons.  Whatever its origin, a nonvanishing
Collins function would be a great gift to the community interested in
the transverse spin structure of the nucleon.  It provides an
unanticipated tool for extracting the nucleon's transversity from DIS
experiments.  The fact that HERMES has seen a robust (2--3\%) asymmetry
with a longitudinally polarized target suggests that they will see a
large asymmetry with a transversely polarized target (unless the
effect is entirely twist three---e.g.,  $h_{L}\gg\delta q$).  This in
turn will lead to the first measurements of the nucleon's transversity
distribution and to new insight into the relativistic spin structure
of confined states of quarks and gluons.

\subsubsection*{Acknowledgements}
This work is supported in part by funds provided by the US
Department of Energy  under cooperative
research agreement \#DF-FC02-94ER40818.


\end{document}